\documentclass[twocolumn]{autart}

\usepackage{color}
\usepackage{xcolor}
\usepackage{graphicx}
\usepackage{caption3}
\usepackage{amsmath}
\usepackage{amssymb}
\usepackage{amsfonts}
\usepackage{mathrsfs}
\usepackage{bm}
\allowdisplaybreaks[1]
\newcounter{cntass}

\begin{document}

\begin{frontmatter}

\title{Robust output regulation of linear system subject to modeled and unmodeled uncertainty\thanksref{footnoteinfo}}

\thanks[footnoteinfo]{This paper was not presented at any IFAC 
meeting. Corresponding author Zhicheng~Zhang.}

\author[tju]{Zhicheng Zhang}\ead{zczhang@tju.edu.cn},
\author[tju]{Zhiqiang Zuo}\ead{zqzuo@tju.edu.cn},
\author[windsor]{Xiang Chen}\ead{xchen@uwindsor.ca},
\author[melbourne]{Ying Tan}\ead{yingt@unimelb.edu.au},
\author[tju]{Yijing Wang}\ead{yjwang@tju.edu.cn}

\address[tju]{School of Electrical and Information Engineering, Tianjin University, Tianjin, 300072, P.~R.~China}
\address[windsor]{Department of Electrical and Computer Engineering, University of Windsor, Windsor, Ontario, N9B 3P4, Canada} 
\address[melbourne]{Department of Mechanical Engineering, University of Melbourne, Parkville, Victoria, 3010, Australia}

\begin{abstract}
	In this paper, a novel robust output regulation control framework is proposed for the system subject to noise, modeled disturbance and unmodeled disturbance to seek tracking performance and robustness simultaneously. The output regulation scheme is utilized in the framework to track the reference in the presence of modeled disturbance, and the effect of unmodeled disturbance is reduced by an $\mathcal{H}_\infty$ compensator. The Kalman filter can be also introduced in the stabilization loop to deal with the white noise. Furthermore, the tracking error in the presence/absence of noise and disturbance is estimated. The effectiveness and performance of our proposed control framework is verified in the numerical example by applying in the Furuta Inverted Pendulum system. 
\end{abstract}


\end{frontmatter}

\section{Introduction}\label{sec_1}

The objective of tracking a reference trajectory in the presence of disturbance is an essential problem running through the development of control theory, and the achievements contribute to the applications including robotic manipulators \cite{Yue2017Model}, quadrotors \cite{Li2019Robust}, autonomous vehicles \cite{Raffo2009Raffo}, \cite{Luan2020Trajectory}, etc.

As disturbances are unavoidable, the exploration of disturbance rejection has attracted comprehensive focus, and lots of outstanding algorithms have been proposed. 
Despite the inherent robustness to tolerate uncertainty in a small range, the investigation of controller design approaches that are dedicated to disturbance rejection is indispensable. The disturbance rejection methods can be roughly divided into two categories, namely, the nonlinear and linear approaches. The former, such as sliding mode control \cite{Rios2019Continuous}, \cite{Wang2021Adaptive}, active disturbance rejection control (ADRC) \cite{Cong2020Active}, etc., play a large enough control torque in suppressing the undesired effect caused by disturbance, and as a result, the zero steady-state error can be acted. 
Compared with targeting at zero static error, the $\mathcal{H}_\infty$ control method, which keeps the objective onto minimize the impact of disturbance on the system, has more practical significance.

Adopting the idea of robust control in the tracking problem, the robust output regulation scheme provides a unified framework for tracking control and disturbance rejection of the system, in which, an external system model is introduced to characterize the disturbance and expected trajectory of the system. The research on robust output regulation scheme has spanned from single systems to multi-agent systems (cf., \cite{Chen2009Global}, \cite{Wang2017Robust}, \cite{Jiang2020Adaptive}, \cite{Liang2021Robust}, \cite{Silani2022Robust}, \cite{Ma2016Cooperative}, \cite{Liu2017Cooperative}). However, this method fails to deal with the model-free disturbance and does not provide more robustness than the inherent robustness of the matrix equation solution, so it is difficult to give a quantitative description of the disturbance rejection performance, which spurs us to investigate the output regulation together with an $\mathcal{H}_\infty$ control loop reducing the effect of disturbance.

This paper proposes a novel control framework consisting of multiple loops to solve an output tracking problem in the presence of modeled/unmodeled disturbance and noise. Both tracking performance and disturbance rejection performance are considered, and the tracking error under the proposed control structure is estimated. 
A distinct feature between our framework and robust output regulation is that the system is subject to the disturbance formulated by an exo-system with known dynamics as described in \cite{huang2004nonlinear} and the unknown disturbance signal simultaneously. Moreover, the upper bound of the influence can be formulated. This problem formulation is applicable to a more general class of dynamic systems.

This paper is organized as follows. The preliminaries and formulation of the main problem are presented in Section~\ref{sec_2}. In Section~\ref{sec_3}, we will design the robust output regulation controller for the system with unmodeled disturbance. Then the extensions to the noisy system and the system subject to both modeled and unmodeled disturbances as well will be presented in Section~\ref{sec_4}. An application example using Furuta Inverted Pendulum is given in Section~\ref{sec_5}, where the comparison between classical output regulation and our proposed design is depicted in order to validate the effectiveness of the proposed design. The paper is concluded in Section~\ref{sec_6}.

\textbf{Notations.} Throughout this paper, all matrices are assumed to have appropriate dimensions, so when it does not cause confusion, some dimensions of the matrices will not be explicitly specified. The transpose of matrix $A$ is expressed as $A^\mathrm{T}$. $\bar{\sigma}(A)$ stands for the maximum singular value of matrix $A$. Let $\|\cdot\|$ be the $2$-norm of a vector. Define the power norm of stochastic signal $w$ as $\|w\|_\mathcal{P}=\sqrt{\lim_{T\rightarrow\infty}\frac{1}{T}\int_0^T\mathbf{E}\left(w(t)^\mathrm{T}w(t)\right)\,dt}$, where $\mathbf{E}(\cdot)$ is expectation. For non-stochastic signal $x$, it reads $\|x\|_\mathcal{P}=\sqrt{\lim_{T\rightarrow\infty}\frac{1}{T}\int_0^Tx(t)^\mathrm{T}x(t)\,dt}$. For transfer functions $\Xi_1=\begin{bmatrix}\Xi_{11}&\Xi_{12}\\\Xi_{21}&\Xi_{22}\end{bmatrix}$, $\Xi_{11}\in\mathbb{C}^{p_1\times q_1}$, $\Xi_{22}\in\mathbb{C}^{p_2\times q_2}$, and $\Xi_2\in\mathbb{C}^{q_2\times p_2}$, the linear fractional transformation is $\mathscr{F}_{\ell}(\Xi_1,\Xi_2):=\Xi_{11}+\Xi_{12}\Xi_2(I-\Xi_{22}\Xi_2)^{-1}\Xi_{21}$, provided that $(I-\Xi_{22}\Xi_2)^{-1}$ exists.

\section{Preliminaries and Formulation of Main Problem}\label{sec_2}

In this paper, we are motivated to investigate the robust output regulation problem for linear systems modeled by
\begin{equation}\label{equ_system_with_w0w1w2}
	\begin{split}		
		\dot{x}&=Ax+B_0w_0+B_1w_1+B_2w_2+Bu, \\
		y& =Cx+D_0w_0+D_1w_1+D_2w_2,
	\end{split}
\end{equation}
where $x\in\mathbb{R}^n$, $u\in\mathbb{R}^m$, and $y\in\mathbb{R}^p$ stand for the state, input and output of the system, respectively. $w_0\in\mathbb{R}^{m_0}$ denotes the white noise signal satisfying $\mathbf{E} \left(w_0(t)\right) = 0$ and $\mathbf{E}\left(w_0(t)w_0^\mathrm{T}(\tau)\right)=\delta(t-\tau)I$. $w_1\in\mathbb{R}^{m_1}$ is the unknown disturbance representing the uncertainty and/or unmodeled dynamics, while $w_2\in\mathbb{R}^{m_2}$ is the measurable disturbance with the dynamics
\begin{equation}\label{equ_disturbance_dynamics}
	\begin{split}		
		\dot{x}_w &=A_wx_w, \\
		w_2 & = C_wx_w.
	\end{split}
\end{equation}

It is assumed that 
\begin{itemize}
	\setlength{\itemindent}{5.5mm}
	\item [(A\refstepcounter{cntass}\thecntass)\label{asm_stabilizable}] 
		$(A,B)$ is stabilizable and $(C,A)$ is detectable;
	\item [(A\refstepcounter{cntass}\thecntass)\label{asm_A_B0_column_rank}]
		$\left[\begin{array}{cc}A-j \omega I & B_0 \\ C & D_0\end{array}\right]$ has full column rank for all $\omega\in\mathbb{R}$; 
	\item [(A\refstepcounter{cntass}\thecntass)\label{asm_D0_full_rank}]
		$D_0D_0^\mathrm{T}>0$;
	\item [(A\refstepcounter{cntass}\thecntass)\label{asm_A_B1_row_rank}]
		$\left[\begin{array}{cc}A-j \omega I & B_1 \\ C & D_1\end{array}\right]$ has full row rank for all $\omega\in\mathbb{R}$; 
	\item [(A\refstepcounter{cntass}\thecntass)\label{asm_D1_full_rank}]
		$\mathrm{rank}\left(D_1\right)=p$. 
	\item [(A\refstepcounter{cntass}\thecntass)\label{asm_A_B_column_rank}]
		$\left[\begin{array}{cc}A-j \omega I & B \\ C & 0\end{array}\right]$ has full column rank for all $\omega\in\mathbb{R}$; 
\end{itemize}

Assumptions (A\ref{asm_stabilizable})--(A\ref{asm_A_B_column_rank}) are basic requirements in LQG and $\mathcal{H}_\infty$ control design \cite{Zhou1998EssentialsRobust}. 

The main objective of this paper is to find a control $u$ such that the output $y$ tracks a reference signal $r\in\mathbb{R}^p$, which is derived from a given model 
\begin{equation}\label{equ_reference_dynamics}
	\begin{split}		
		\dot{x}_r&=A_rx_r,\quad x_r(0)=x_{r0}, \\
		r& =C_rx_r, 
	\end{split}
\end{equation}
that is, the tracking error $e=y-C_rx_r$ satisfies $\|e\|_{\mathcal P}$ is bounded in the presence of noise $w_0$, bounded uncertain disturbance $w_1$ and modeled disturbance $w_2$, and, in particular, $\lim_{t\rightarrow\infty}\|e(t)\|_{\mathcal P}=0$ for $w_0=0$, $w_1=0$ and $w_2=0$. 

As commonly appeared in the theory of output regulation, the following assumptions hold in subsequent discussions.
\begin{itemize}
	\setlength{\itemindent}{5.5mm}
	\item [(A\refstepcounter{cntass}\thecntass)\label{asm_OR_detectable_ArAw}]
		$(C_r,A_r)$ and $(C_w, A_w)$ are both detectable; 
	\item [(A\refstepcounter{cntass}\thecntass)\label{asm_OR_eigenvalue_ArAw}]
		$A_r$ and $A_w$ have no eigenvalues with negative real parts;
	\item [(A\refstepcounter{cntass}\thecntass)\label{asm_OR_solvable_ArAw}]
		$\begin{bmatrix}A-\lambda I & B \\ C & 0\end{bmatrix}$ has full row rank for all eigenvalues $\lambda$ of $A_r$ and $A_w$;
\end{itemize}

In order to present our control design idea clearly, the tracking control of the noise- and disturbance-free system is first discussed, and we have the following theorem. 

\begin{thm}\label{thm_output_regulation}\cite{huang2004nonlinear}
	For system 
	\begin{equation*} 
		\begin{split}		
			\dot{x}&=Ax+Bu, \\
			y& =Cx,
		\end{split}
	\end{equation*}
	satisfying Assumption (A\ref{asm_stabilizable}), and reference dynamics (\ref{equ_reference_dynamics}) satisfying Assumptions (A\ref{asm_OR_detectable_ArAw})--(A\ref{asm_OR_solvable_ArAw}), the output regulation problem is solved by controller 
	\begin{align}
		\dot{\hat{x}} &= A\hat{x}+Bu - L_t\left(y-C\hat{x}\right), \label{equ_OR_controller_hat_x}\\
		\dot{\hat{x}}_r &= A_r\hat{x}_r-L_r(r-C_r\hat{x}_r), \label{equ_OR_controller_hat_xr}\\
		u=u_t&=K_t\hat{x}+K_r\hat{x}_r, \label{equ_OR_controller_u}
	\end{align}
	where $K_t$, $L_r$ and $L_t$ are gains such that $(A+BK_t)$, $(A+L_tC)$ and $(A_r+L_rC_r)$ are Hurwitz, and $K_r$ satisfies
	\begin{align}
		X_tA_r &= AX_t+BK_r+BK_tX_t, \label{equ_OR_condition_1}\\
		0&=CX_t-C_r,	\label{equ_OR_condition_2}
	\end{align}
	for an appropriate matrix $X_t$. 
\end{thm}

For systems with uncertainties or disturbances, the inherent robustness of the system is not sufficient to maintain tracking performance. Therefore, the robust control method is considered. 

\begin{figure}[!htp]
	\centering
	\includegraphics[width=8.3cm]{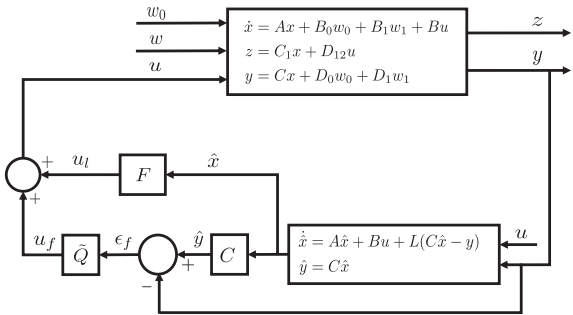}
	\caption{Complementary Control Structure}
	\label{fig_LQG_with_Hinfty}
\end{figure}

While traditional robust methods, such as ${\mathcal H}_\infty$ control\cite{Zhou1996RobustOptimal} or mixed ${\mathcal H_2}/{\mathcal H}_\infty$ control \cite{Zhou1994MixedH2Hinfty}, could be applied to address this robust output regulation problem, it is also well-known that these methods could be quite conservative which may not provide good tracking performance and robustness simultaneously. Noteworthy, a new development is reported in \cite{chen2019revisit} which presents a complementary control structure derived based on Youla parameterization of all stabilizing controllers as shown in Fig.~\ref{fig_LQG_with_Hinfty}, where this new structure effectively combines an $LQG$ control and an ${\mathcal H}_\infty$ control to achieve non-compromised optimal performance and robustness, as summarized in Theorem \ref{thm_LQG_with_Hinfty}. 

\begin{thm}\label{thm_LQG_with_Hinfty}\cite{chen2019revisit}
	For system 
	\begin{align*}
		\dot{x}&=A x+B_0 w_0+B_1 w_1+B u, \\
		z&=C_1 x+D_{12} u, \\
		y&=C x+D_0 w_0+D_1 w_1,
	\end{align*}
	satisfying Assumptions (A\ref{asm_stabilizable})--(A\ref{asm_D1_full_rank}), the combined LQG/$\mathcal{H}_\infty$ controller can be designed as $u=u_l+u_\infty$ with the LQG controller 
	\begin{equation*}
		\begin{split}
			\dot{\hat{x}} &=\left(A+L C\right) \hat{x}+B u-L y, \\
			u_l &=F \hat{x},
		\end{split}
	\end{equation*}
	and $\mathcal{H}_\infty$ controller $\tilde{Q}$, which is of the form 
	\begin{equation*}
		\begin{split}
			\dot{x}_\infty&=A_\infty x_\infty+B_\infty \epsilon_\infty, \\
			u_\infty&=F_\infty x_\infty,
		\end{split}
	\end{equation*}
	using residual signal $\epsilon_\infty=C\hat{x}-y$, where 
	\begin{align*}
		F &=-R_1^{-1}\left(D_{12}^\mathrm{T} C_1+B^\mathrm{T} P_{10}\right), \\
		L &=-\left(B_0 D_0^\mathrm{T}+P_{20} C^\mathrm{T}\right) R_0^{-1}, \\
		A_\infty&=\bar{A}+\gamma^{-2} \bar{B}_1 \bar{B}_1^\mathrm{T} \bar{P}_1+\bar{B}_2 F_\infty \\
		&\quad -B_\infty\left(\bar{C}_2-\gamma^{-2} D_1 \bar{B}_1^\mathrm{T} \bar{P}_1\right), \\
		B_\infty&=-\left(I-\gamma^{-2} \bar{P}_2 \bar{P}_1\right)^{-1} L_\infty, \\
		F_\infty&=-\left[F+R_1^{-1}\left(B^\mathrm{T} P_1+D_{12}^\mathrm{T} C_1\right),\; -F\right], \\
		L_\infty&=\left[\begin{array}{c}
		\left(P_2 C^\mathrm{T}+B_1 D_1^\mathrm{T}\right) R_2^{-1} \\
		\left(P_2 C^\mathrm{T}+B_1 D_1^\mathrm{T}\right) R_2^{-1}+L
		\end{array}\right], 
	\end{align*}
	and $P_{10} \geq 0$, $P_{20} \geq 0$, $P_1\geq 0$, $P_2\geq 0$, satisfying $\rho(P_1P_2)<\gamma^2$, solve
	\begin{align*}
		&P_{10} A_1+A_1^\mathrm{T} P_{10}-P_{10} B R_1^{-1} B^\mathrm{T} P_{10}+Q_1=0, \\
		&P_{20} A_0^\mathrm{T}+A_0 P_{20}-P_{20} C^\mathrm{T} R_0^{-1} C P_{20}+Q_0=0, \\
		&P_1A_1+A_1^\mathrm{T} P_1+P_1\left(\frac{B_1 B_1^\mathrm{T}}{\gamma^2}-B R_1^{-1} B^\mathrm{T}\right) P_1+Q_1=0,\\
		&P_2 A_2^\mathrm{T}+A_2 P_2+P_2\left(\frac{C_1^\mathrm{T} C_1}{\gamma^2}-C^\mathrm{T} R_2^{-1} C\right) P_2+Q_2=0,\\
		&\bar{P}_1=\left[\begin{array}{cc}
			P_1 & 0 \\
			0 & 0
			\end{array}\right], \; \bar{P}_2=\left[\begin{array}{ll}
			P_2 & P_2 \\
			P_2 & P_2
			\end{array}\right], 
	\end{align*}
	with 
	\begin{align*}
		&\bar{A}=\left[\begin{array}{cc}
			A+B F & -B F \\
			0 & A+L C
			\end{array}\right], \\ 
		&\bar{B}_1=\left[\begin{array}{c}
			B_1 \\
			B_1+L D_1
			\end{array}\right], \;\bar{B}_2=\left[\begin{array}{c}
			B \\
			0
			\end{array}\right], \\
		&\bar{C}_2=[0,\, -C], \\
		&A_0=A-B_0 D_0^\mathrm{T} R_0^{-1} C,\; A_1=A-B R_1^{-1} D_{12}^\mathrm{T} C_1,\\
		& A_2=A-B_1 D_1^\mathrm{T} R_2^{-1} C,  \\
		&Q_0=B_0\left(I-D_0^\mathrm{T} R_0^{-1} D_0\right) B_0^\mathrm{T},\\
		&Q_1=C_1^\mathrm{T}\left(I-D_{12} R_1^{-1} D_{12}^\mathrm{T}\right) C_1, \\
		& Q_2=B_1\left(I-D_1^\mathrm{T} R_2^{-1} D_1\right) B_1^\mathrm{T},\\
		& R_1 = D_{12}^\mathrm{T}D_{12},\quad R_2=D_1D_1^\mathrm{T}.
	\end{align*}
	Furthermore, $\|z\|_\mathcal{P}/\|w_1\|_\mathcal{P}<\gamma$ with $\gamma>0$. 
\end{thm}

In the following, a novel control strategy is proposed for the system in the presence of $w_1$ is discussed first, where the main advantages of our proposed framework are manifested. The situation with noise and modeled disturbance will then be analyzed respectively. At last, we extend the control framework to the system that is not fully detectable. 

\section{Design of Robust Output Regulation Control}\label{sec_3}

\begin{figure}[!htp]
	\centering
	\includegraphics[width=8.3cm]{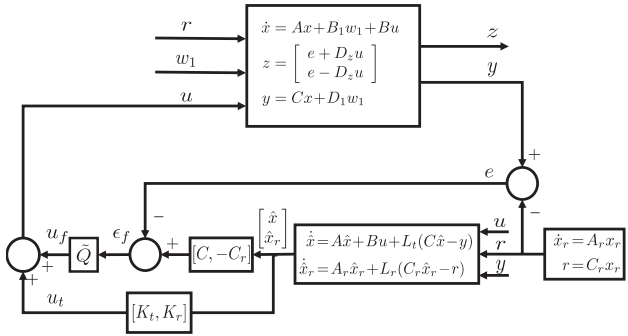}
	\caption{Robust Output Regulation Control Structure}
	\label{fig_novel_structure}
\end{figure}

Motivated by Theorems \ref{thm_output_regulation} and \ref{thm_LQG_with_Hinfty}, we present a novel robust output regulation control strategy for disturbed system, as shown in Fig.~\ref{fig_novel_structure}, whose dynamics is characterized as
\begin{equation*} 
	\begin{split}		
		\dot{x}&=Ax+B_1w_1+Bu, \\
		y&=Cx+D_1w_1.
	\end{split}
\end{equation*}

In this control strategy, both output regulation performance and robustness are considered and handled by the output regulation control $(K_r, L_r)$ and $\tilde Q$ respectively. The measured and estimated tracking errors are used to find the residual signal $\epsilon_f$ reflecting the impact of disturbance $w_1$. 

From Theorem~\ref{thm_output_regulation}, the output regulation scheme can be designed as (\ref{equ_OR_controller_hat_x})--(\ref{equ_OR_controller_u}), which will give a control singal $u_t$ to perfectly track the reference $r$ in the absence of unmodeled disturbance $w_1$. 

In order to evaluate the tracking performance, a vector variable $z=\left[\begin{array}{c}e+D_zu\\e-D_zu\end{array}\right]\in\mathbb{R}^{2p}$ is introduced into the system, namely,
\begin{equation*}
	\begin{split}		
		\dot{x}&=Ax+B_1w_1+Bu, \\
        z&=\left[\begin{array}{c}e+D_zu\\e-D_zu\end{array}\right],\\
		y& =Cx+D_1w_1,
	\end{split}
\end{equation*}
where $D_z$ is a matrix with $\mathrm{rank} D_z=m$. 

To design the controller $\tilde{Q}$, we first reformulate the system. Let
$$x_f=\begin{bmatrix}
	\hat{x}-X_t x_r \\
	x-\hat{x}
\end{bmatrix}$$ 
with $X_t$ being defined in (\ref{equ_OR_condition_1}) and (\ref{equ_OR_condition_2}), and $\overline{w}_1=x_r-\hat{x}_r$, then the augmented system together with the output regulation scheme is of the form 
\begin{equation}\label{equ_system_for_Hinfty}
	\begin{aligned}
	\dot{x}_f &= \begin{bmatrix}
			A+BK_t & -L_tC \\
			0 & A+L_tC
		\end{bmatrix}x_f + \begin{bmatrix}
			-BK_r \\ 0
		\end{bmatrix}\overline{w}_1 \\
		&\quad  + \begin{bmatrix}
			-L_tD_1 \\ B_1+L_tD_1
		\end{bmatrix}w_1 + \begin{bmatrix}
			B \\ 0
		\end{bmatrix}u_f, \\
	z &= \begin{bmatrix}
			C & C \\
			C & C
		\end{bmatrix}x_f + \begin{bmatrix}
			D_1 \\
			D_1
		\end{bmatrix}w_1 + \begin{bmatrix}
			D_z \\ -D_z
		\end{bmatrix} u_f, \\
	\epsilon_f&=\begin{bmatrix}
			0 & -C
		\end{bmatrix}x_f + C_r\overline{w}_1 - D_1w_1, 	
	\end{aligned}
\end{equation}
where residual signal $\epsilon_f=C \hat{x}-C_r\hat{x}_r-e$ describes the difference between the nominal model and the real system. 

Since $\lim_{t\rightarrow\infty}\overline{w}_1=0$, we can ignore $\overline{w}_1$ and keep focus on the unknown disturbance $w_1$. By introducing linear transformation
$$\begin{array}{ll}
	u_f=V_2^\mathrm{T}\Sigma_2^{-1}{\tilde{u}_f}, & w_1=V_1^\mathrm{T}{\tilde{w}}, \\
	\epsilon_f=U_1^\mathrm{T}\Sigma_1{\tilde{y}_f}, & z=U_2^\mathrm{T}{\tilde{z}},
\end{array}$$
where the matrices $U_1$, $U_2$, $V_1$, $V_2$, $\Sigma_1$ and $\Sigma_2$ satisfy
\begin{align*}
	-D_1&=U_1^\mathrm{T}\begin{bmatrix}
		0 & \Sigma_1
	\end{bmatrix}V_1, \\
	\begin{bmatrix}
		D_z \\ -D_z
	\end{bmatrix}&=U_2^\mathrm{T}\begin{bmatrix}
		0 \\ \Sigma_2
	\end{bmatrix}V_2,
\end{align*}
with $\mathrm{rank}(\Sigma_1)=p$, $\mathrm{rank}(\Sigma_2)=m$. Under Assumption (A\ref{asm_D1_full_rank}), the system (\ref{equ_system_for_Hinfty}) can be rewritten as
\begin{equation}\label{equ_system_for_Hinfty_2}
	\begin{aligned}
		\dot{x}_f &= A_f x_f + B_{1f} \tilde{w} + B_{2f}\tilde{u}_f, \\
		\tilde{z}&= C_{1f} x_f + D_{11f} \tilde{w} + D_{12f}\tilde{u}_f, \\
		\tilde{y}_f&= C_{2f} x_f + D_{21f} \tilde{w},
	\end{aligned}
\end{equation}
where
\begin{gather*}
	A_f = \begin{bmatrix}
		A+BK_t & -L_tC \\
		0 & A+L_tC
	\end{bmatrix}, \\
	B_{1f} = \begin{bmatrix}
		-L_tD_1 \\ B_1+L_tD_1
	\end{bmatrix}V_1^\mathrm{T}, \;\;	B_{2f} = \begin{bmatrix}
		B \\ 0
	\end{bmatrix}V_2^\mathrm{T}\Sigma_2^{-1}, \\
	C_{1f} = U_2\begin{bmatrix}C & C \\ C & C\end{bmatrix}, \;\;	C_{2f} = \Sigma_1^{-1}U_1\begin{bmatrix}0 & -C\end{bmatrix}, \\
	D_{12f} = \begin{bmatrix}0\\I_m\end{bmatrix}, \;\;	D_{21f} = \begin{bmatrix}0&I_p\end{bmatrix},
\end{gather*}
and $D_{11f}=U_2\left[D_1^\mathrm{T},\ D_1^\mathrm{T}\right]^\mathrm{T}V_1^\mathrm{T}$ can be decomposed into
$$D_{11f} = \begin{bmatrix}
			0 & D_{1112} \\
			0 & D_{1122}
		\end{bmatrix},\quad D_{1122}\in\mathbb{R}^{m\times p}.$$

Given positive scalar $\gamma$, denote 
\begin{align*}
	\Gamma &= \left(D_{1112}^\mathrm{T}D_{1112}-\gamma^2I_p\right)^{-1}, \\
	\mathcal{D} &= D_{1112}^\mathrm{T}D_{1112}+D_{1122}^\mathrm{T}D_{1122}, 
\end{align*}
and symmetric matrices
\begin{align*}
	\mathcal{R} &= \begin{bmatrix}
		-\gamma^{-2}I_{m_1-p} & 0 & 0 \\
		0 & \Gamma & -\Gamma D_{1122}^\mathrm{T} \\
		0 & -D_{1122}\Gamma & I_{m}+D_{1122}\Gamma D_{1122}^\mathrm{T}
	\end{bmatrix}, \\
	\tilde{\mathcal{R}} &= \begin{bmatrix}
			-\gamma^{-2}I_{2p-m} & 0 & \gamma^{-2}D_{1112} \\
			0 & -\gamma^{-2}I_{m} & \gamma^{-2}D_{1122} \\
			\gamma^{-2}D_{1112}^\mathrm{T} & \gamma^{-2}D_{1122}^\mathrm{T} & I_p-\gamma^{-2}\mathcal{D}
		\end{bmatrix}.
\end{align*}
Then with the controller gain
\begin{align*}
	K_f &\ =-\mathcal{R}\Big(\left[D_{11f},\,D_{12f}\right]^\mathrm{T}  C_{1f}+\left[B_{1f},\ B_{2f}\right]^\mathrm{T} X_f\Big) \\
		&:=\left[K_{11f}^\mathrm{T},\,K_{12f}^\mathrm{T},\,K_{2f}^\mathrm{T}\right]^\mathrm{T}, \\
	L_f &\ =-\Big(B_{1f} \left[D_{11f}^\mathrm{T},\,D_{21f}^\mathrm{T}\right]+Y_f \left[C_{1f}^\mathrm{T},\ C_{2f}^\mathrm{T}\right]\Big) \tilde{\mathcal{R}} \\
		&:=\left[L_{11f},\,L_{12f},\,L_{2f}\right],
\end{align*}
where $K_{11f}\in\mathbb{R}^{(m_1-p)\times 2n}$, $K_{12f}\in\mathbb{R}^{p\times 2n}$, $K_{2f}\in\mathbb{R}^{m\times 2n}$, $L_{11f}\in\mathbb{R}^{2n\times (2p-m)}$, $L_{12f}\in\mathbb{R}^{2n\times m}$, $L_{2f}\in\mathbb{R}^{2n\times p}$, $X_f$ and $Y_f$ are, respectively, the solutions to the Ricatti equations
\begin{align*}
	H_{11}^\mathrm{T}X_f+X_fH_{11}+X_fH_{12}X_f + H_{21}&=0, \\
	J_{11}^\mathrm{T}Y_f+Y_fJ_{11}+Y_fJ_{12}Y_f + J_{21}&=0,
\end{align*}
with
\begin{align*}
	H_{11} &= A_f+\left[B_{1f},\ B_{2f}\right]\begin{bmatrix}0_{(m_1-p)\times(2p-m_2)}&0\\-\Gamma D_{1112}^\mathrm{T} & 0 \\ D_{1122}\Gamma D_{1112}^\mathrm{T} & -I_{m}\end{bmatrix}C_{1f}, \\
	H_{12} &= -\left[B_{1f},\ B_{2f}\right]\mathcal{R}\left[B_{1f},\ B_{2f}\right]^\mathrm{T}, \\
	H_{21} &= C_{1f}^\mathrm{T}\begin{bmatrix}I_{2p-m}-D_{1112}\Gamma D_{1112}^\mathrm{T} & 0 \\ 0 & 0_{m} \end{bmatrix}C_{1f}, \\
	J_{11} &= A_f^\mathrm{T}-\begin{bmatrix}0_{n\times (m_1-p)} & C_{2f}^\mathrm{T}\end{bmatrix} B_{1f}^\mathrm{T}, \\
	J_{12} &= -\begin{bmatrix}C_{1f}\\C_{2f}\end{bmatrix}^\mathrm{T}\tilde{\mathcal{R}}\begin{bmatrix}C_{1f}\\C_{2f}\end{bmatrix}, \\
	J_{21} &= B_{1f}\begin{bmatrix}I_{m_1-p} & 0 \\ 0 & 0_p\end{bmatrix} B_{1f}^\mathrm{T}, \\
\end{align*}
the controller can be designed as
\begin{equation}\label{equ_tildeQ_controller}
	\mathscr{F}_{\ell}\left(\mathbf{K}_f, \Xi\right),
\end{equation}
where $\Xi$ is any dynamics with input $\Xi_u$ and output $\Xi_y$ satisfying that $\Xi \in R H_{\infty}$, $\|\Xi\|_{\infty}<\gamma$. $\mathbf{K}_f$ can be formulated as
\begin{equation*}
	\begin{aligned}
		\dot{\hat{x}}_f &= \hat{A} \hat{x}_f + \hat{B}_1\Sigma_1^{-1}U_1 \epsilon_f + \hat{B}_2 \Xi_y, \\
		u_f &= V_2^\mathrm{T}\Sigma_2^{-1}\hat{C}_1 \hat{x}_f + V_2^\mathrm{T}\Sigma_2^{-1}\hat{D}_{11}\Sigma_1^{-1}U_1 \epsilon_f \\
		&\quad + V_2^\mathrm{T}\Sigma_2^{-1}\hat{D}_{12} \Xi_y, \\
		\Xi_u &= \hat{C}_2 \hat{x}_f + \hat{D}_{21}\Sigma_1^{-1}U_1 \epsilon_f,
	\end{aligned}
\end{equation*}
with $\hat{D}_{11}=-D_{1122}$, $\hat{D}_{12} \in \mathbb{R}^{m \times m}$ and $\hat{D}_{21} \in \mathbb{R}^{p \times p}$ being any matrices (e.g. Cholesky factors) satisfying $\hat{D}_{12} \hat{D}_{12}^\mathrm{T}=I$, $\hat{D}_{21}^\mathrm{T} \hat{D}_{21}=I-\gamma^{-2}D_{1112}^\mathrm{T}D_{1112}$, $\Xi_u$ and $\Xi_y$ being respectively the input and output variables of $\Xi$, and
\begin{align*}
	\hat{B}_2&=\left(B_{2f}+L_{12f}\right) \hat{D}_{12}, \\
	\hat{C}_2&=-\hat{D}_{21}\left(C_{2f}+K_{12f}\right) Z, \\
	\hat{B}_1&=-L_{2f}+\hat{B}_2 \hat{D}_{12}^{-1} \hat{D}_{11}, \\
	\hat{C}_1&=K_{2f} Z+\hat{D}_{11} \hat{D}_{21}^{-1} \hat{C}_2, \\
	\hat{A}&=A_f+L_f\left[C_{1f}^\mathrm{T},\,C_{2f}^\mathrm{T}\right]^\mathrm{T}+\hat{B}_2 \hat{D}_{12}^{-1} \hat{C}_1, \\
	Z&=\left(I-\gamma^{-2} Y_f X_f\right)^{-1}.
\end{align*}

The following theorem indicates the stability and performance of the closed-loop system.

\begin{thm}\label{thm_tildeQ_control}
	For system (\ref{equ_system_for_Hinfty}) satisfying Assumptions (A\ref{asm_stabilizable}), (A\ref{asm_A_B1_row_rank})--(A\ref{asm_A_B_column_rank}) and the reference dynamics (\ref{equ_reference_dynamics}) satisfying Assumptions (A\ref{asm_OR_detectable_ArAw})--(A\ref{asm_OR_solvable_ArAw}), if for given positive constant $\gamma>\bar{\sigma}(D_{1112})$, there exist $X_f$ and $Y_f$ such that $\rho(X_fY_f)<\gamma^2$, then the controller (\ref{equ_tildeQ_controller}) stabilizes the system (\ref{equ_system_for_Hinfty}). Furthermore, the tracking error satisfies
	\begin{equation}\label{equ_tracking_error_with_w1}
		\|e\|_\mathcal{P}< \gamma \|w_1\|_\mathcal{P}.
	\end{equation}
\end{thm}

\begin{pf}
	First of all, we will prove that the conditions for $\mathcal{H}_\infty$ control hold. Under the controller (\ref{equ_OR_controller_hat_x})--(\ref{equ_OR_controller_u}), $A_f$ is Hurwitz, which implies that the system is stabilizable and detectable. Since $\mathrm{rank}\left(D_z\right)=m$, one has
	\begin{equation*}
		\begin{bmatrix}
			A+BK_t-j\omega I  & -L_tC & B \\
			0 & A+L_tC-j\omega I  & 0 \\
			C & C & D_z \\
			C & C & -D_z
		\end{bmatrix}
	\end{equation*}
	has full column rank for any $\omega\in\mathbb{R}$. Similarly, assumption (A\ref{asm_A_B1_row_rank}) ensures
	\begin{equation*}
		\begin{bmatrix}
			A+BK_t-j\omega I  & -L_tC & -L_tD_1 \\
			0 & A+L_tC-j\omega I  & B_1+L_tD_1 \\
			0 & -C & -D_1
		\end{bmatrix}
	\end{equation*}
	has full row rank.
	
	Then to analyze the stability of the closed-loop system, denote
	\begin{align*}
		R&= \begin{bmatrix}
			D_{11f}^\mathrm{T} \\ D_{12f}^\mathrm{T}
		\end{bmatrix}\begin{bmatrix}
			D_{11f} & D_{12f}
		\end{bmatrix}-\begin{bmatrix}
			\gamma^2I_{m_1} & 0 \\ 0 & 0
		\end{bmatrix}, \\
		\tilde{R}&=\begin{bmatrix}
			D_{11f} \\ D_{21f}
		\end{bmatrix}\begin{bmatrix}
			D_{11f}^\mathrm{T} & D_{21f}^\mathrm{T}
		\end{bmatrix}-\begin{bmatrix}
			\gamma^2I_{2p} & 0 \\ 0 & 0
		\end{bmatrix},
	\end{align*}
	and it is clear that $R\mathcal{R}=\mathcal{R}R=I$, $\tilde{R}\tilde{\mathcal{R}}=\tilde{\mathcal{R}}\tilde{R}=I$. According to the main result in \cite{glover1988state}, $K_f$ and $L_f$ with respect to $X_f$ and $Y_f$ are proper controller gains to solve the $\mathcal{H}_\infty$ control problem. Thereby, the controller $\mathscr{F}_{\ell}\left(\widetilde{\mathbf{K}}_f, \Xi\right)$ with $\widetilde{\mathbf{K}}_f$ being designed as
	\begin{equation*}
		\begin{aligned}
		\dot{\hat{x}}_f &= \hat{A} \hat{x}_f+ \hat{B}_1 \tilde{y}_f + \hat{B}_2 \Xi_y, \\
		\tilde{u}_f &= \hat{C}_1 \hat{x}_f + \hat{D}_{11} \tilde{y}_f + \hat{D}_{12} \Xi_y, \\
		\Xi_u &= \hat{C}_2 \hat{x}_f + \hat{D}_{21} \tilde{y}_f,
		\end{aligned}
	\end{equation*}
	is an admissible stabilizer for the system (\ref{equ_system_for_Hinfty_2}), and equivalently, the system (\ref{equ_system_for_Hinfty}) complete with controller (\ref{equ_tildeQ_controller}) achieves $\frac{\|z\|_\mathcal{P}}{\|w_1\|_\mathcal{P}}< \gamma$ if $\frac{\|\tilde{z}\|_\mathcal{P}}{\|\tilde{w}\|_\mathcal{P}}=\frac{\|z\|_\mathcal{P}}{\|w_1\|_\mathcal{P}}$ is noticed.

	To estimate the effect of the disturbance on the output, consider that
	$$
		\|z\|_\mathcal{P}^2 =\left\|\begin{bmatrix}
			e+D_zu \\ e-D_zu
		\end{bmatrix}\right\|_\mathcal{P}^2 = \|e\|_\mathcal{P}^2+\left\|D_zu\right\|_\mathcal{P}^2 \\
		\geq \|e\|_\mathcal{P}^2.
	$$
	Then (\ref{equ_tracking_error_with_w1}) is obtained, which completes the proof.
\end{pf}

Theorem \ref{thm_tildeQ_control} provides an approach to suppress the effect of the uncertainty/disturbance using the residual signal. Combined with Theorem \ref{thm_output_regulation}, the robust tracking objective can be fulfilled by focus on the tracking performance first to get $u_t$ defined in \ref{equ_OR_controller_u}, and then, if the system deviates from the nominal model, the $\mathcal{H}_\infty$ control $\tilde{Q}$ is designed to get $u_f$. Then one gets the complemental control 
\begin{equation}\label{equ_control_utuf}
	u=u_t+u_f.
\end{equation}
If there is no uncertainty/disturbance in the system, $\tilde{Q}$ will not operate and the tracking performance appears.

\begin{rem}
	The performance output $z$ in this paper includes direct disturbances term $D_1w_1$, $\gamma>\bar{\sigma}(D_{1112})$ is hence a sufficient condition for $\mathcal{H}_\infty$ control.
	Specially, when $m=2p$, the condition degenerates into $\gamma>0$, which is in coincidence with that in \cite{chen2019revisit}.
\end{rem}

\section{Robust Output Regulation for Linear Systems with Noise And Modeled/Unmodeled Disturbance}\label{sec_4}

In this section, the robust output regulation control for linear systems subject to, simultaneously, white noise, model disturbance and unmodeled disturbance is discussed. 

\subsection{Linear System with White Noise}

\begin{figure}[!htp]
	\centering
	\includegraphics[width=8.3cm]{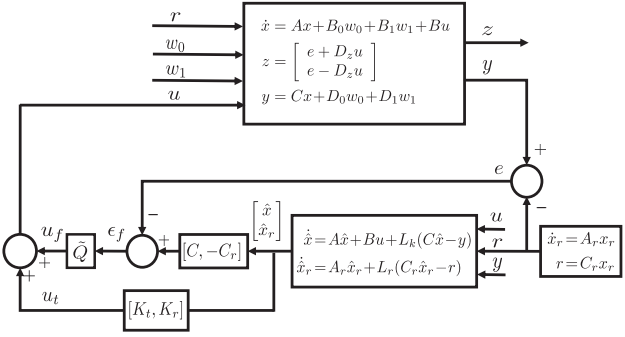}
	\caption{Control Structure for System with Noise}
	\label{fig_novel_structure_with_w0}
\end{figure}

The white noise is addressed in this subsection, for which, the control structure is depicted in Fig.~\ref{fig_novel_structure_with_w0}.

For the noisy system 
\begin{equation}\label{equ_system_with_w0w1}
	\begin{split}		
		\dot{x}&=Ax+B_0w_0+B_1w_1+Bu, \\
		y&=Cx+D_0w_0+D_1w_1,
	\end{split}
\end{equation}
an observer acting as an Kalman filter \cite{Zhou1996RobustOptimal} is employed in (\ref{equ_OR_controller_hat_x}), which is redesigned as 
\begin{equation} \label{equ_OR_controller_Kalman}
	\dot{\hat{x}} = A\hat{x}+Bu_t - L_k\left(y-C\hat{x}\right),
\end{equation}
where $L_k=-(B_0D_0^\mathrm{T}+P_1C^\mathrm{T})R_0^{-1}$ with $R_0=D_0D_0^\mathrm{T}$, $P_1>0$ is the solution to the Ricatti equation
\begin{equation}\label{equ_ARE_Kalman}
	\begin{split}	
		&(A-B_0D_0^\mathrm{T}R_0^{-1}C)P_1+P_1(A-B_0D_0^\mathrm{T}R_0^{-1}C)^\mathrm{T} \\
		& -P_1C^\mathrm{T}R_0^{-1}CP_1 + B_0(I-D_0^\mathrm{T}R_0^{-1}D_0)B_0=0.
	\end{split}	
\end{equation}
Denote $P_2>0$ meets $(A+BK_t)P_2+P_2(A+BK_t)^\mathrm{T}+L_kD_0D_0^\mathrm{T}L_k^\mathrm{T}=0$, then we have the following result. 

\begin{thm}\label{thm_control_with_w0w1}
	Given reference dynamics (\ref{equ_reference_dynamics}) satisfying Assumptions (A\ref{asm_OR_detectable_ArAw})--(A\ref{asm_OR_solvable_ArAw}). Consider system (\ref{equ_system_with_w0w1}) satisfying Assumptions (A\ref{asm_stabilizable})--(A\ref{asm_A_B_column_rank}), under controller (\ref{equ_OR_controller_hat_xr}), (\ref{equ_tildeQ_controller}), (\ref{equ_OR_controller_Kalman}) and (\ref{equ_control_utuf}). The tracking error $e$ satisfies 
	\begin{equation}\label{equ_tracking_error_with_w0w1}
		\|e\|_\mathcal{P} = \sqrt{\mathrm{tr}\left(C(P_1+P_2)C^\mathrm{T} + D_0D_0^\mathrm{T}\right) + \gamma^2 \|w_1\|_\mathcal{P}^2}.
	\end{equation}
	In particular, for the case where $w_0=0$, (\ref{equ_tracking_error_with_w1}) holds.
\end{thm}

\begin{pf}
	First, we will show that $L_k$ is a proper gain for the output regulation scheme. Rewrite (\ref{equ_ARE_Kalman}) as 
	\begin{align*}
		&(A+L_kC)P_1+P_1(A+L_kC)^\mathrm{T} \\
		&\quad +(B+L_kD_0)(B+L_kD_0)^\mathrm{T}=0,
	\end{align*}
	which means $A+L_kC$ is Hurwitz and $L_k$ meets the requirement of output regulation design.

	Second, the baseline performance is discussed, and the impact of noise $w_0$ on tracking error is deduced in the absence of $w_1$. Let 
	$$\tilde{x}_t=\begin{bmatrix}
		\hat{x}-X_tx_r \\
		x-\hat{x}
	\end{bmatrix},$$
	then, one has   
	\begin{equation}
		\begin{split}
			\dot{\tilde{x}}_t &= \tilde{A}_t\tilde{x}_t + \tilde{B}_rw_r+\tilde{B}_0w_0, \\
			e &= \tilde{C}_t\tilde{x}_t + D_0w_0,
		\end{split}		
	\end{equation}
	with $\overline{w}_1=\hat{x}_r-x_r$, and 
	\begin{align*}
		&\tilde{A}_t= \begin{bmatrix}
			A+BK_t & -L_kC \\
			0 & A+L_kC 
		\end{bmatrix}, \\
		&  \tilde{B}_0= \begin{bmatrix}
			-L_kD_0 \\ B_0+L_kD_0 
		\end{bmatrix}, \; \tilde{B}_r= \begin{bmatrix}
			BK_r \\ 0
		\end{bmatrix}, \\
		& \tilde{C}_t=\begin{bmatrix}C & C\end{bmatrix}.
	\end{align*}
	Since $\overline{w}_1$ vanishes exponentially, $\tilde{B}_r \overline{w}_1$ can be ignored. The tracking error is measured by 
	$$e=\tilde{C}_t\int_0^t\Phi(t,\tau)\tilde{B}_0w_0(\tau)\,d\tau+D_0w_0,$$
	where $\Phi(t,\tau)=e^{-\tilde{A}_t(t-\tau)}$. Then one has
	\begin{align*}
		\|e\|_\mathcal{P}^2 &= \lim_{T\rightarrow\infty}\frac{1}{T}\mathbf{E}\bigg(\int_0^T\bigg(\int_0^t\int_0^tw_0(\tau)^\mathrm{T}\tilde{B}_0^\mathrm{T}\Phi(t,\tau)^\mathrm{T}\tilde{C}_t^\mathrm{T}  \\
		&\quad \times \tilde{C}_t\Phi(t,s)\tilde{B}_0w_0(s)\,ds\,d\tau +w_0^\mathrm{T}D_0^\mathrm{T}D_0w_0 \\
		&\quad +2w_0^\mathrm{T}D_0^\mathrm{T}\tilde{C}_t\int_0^t\Phi(t,\tau)\tilde{B}_0w_0(\tau)\,d\tau\bigg)dt\bigg)  \\
		&= \lim_{T\rightarrow\infty}\frac{1}{T}\mathrm{tr}\bigg(\int_0^T\bigg(\int_0^t\int_0^t\tilde{C}_t\Phi(t,s)\tilde{B}_0 \\
		&\quad \times \mathbf{E}(w_0(s)w_0(\tau)^\mathrm{T})\tilde{B}_0^\mathrm{T}\Phi(t,\tau)^\mathrm{T}\tilde{C}_t^\mathrm{T}\,ds\,d\tau \\
			& \quad +2\int_0^t\Phi(t,\tau)\tilde{B}_0\mathbf{E}\left(w_0(\tau)w_0(t)^\mathrm{T}\right)D_0^\mathrm{T}\tilde{C}_t\,d\tau \\
			&\quad +D_0\mathbf{E}(w_0(t)w_0(t)^\mathrm{T})D_0^\mathrm{T} \bigg)dt\bigg) \\
		&= \lim_{T\rightarrow\infty}\frac{1}{T}\bigg(\mathrm{tr}\bigg(\int_0^T\int_0^t\tilde{C}_t\Phi(t,s)\tilde{B}_0 \\
		&\quad \times \tilde{B}_0^\mathrm{T}\Phi(t,s)^\mathrm{T}\tilde{C}_t^\mathrm{T}\,ds\,dt\bigg) +T\mathrm{tr}\left(D_0D_0^\mathrm{T}\right)\bigg) \\
		&= \lim_{T\rightarrow\infty}\frac{1}{T}\mathrm{tr}\left(\int_0^T\tilde{C}_tY\tilde{C}_t^\mathrm{T}dt\right) +\mathrm{tr}\left(D_0D_0^\mathrm{T} \right),
	\end{align*}
	where $Y=\int_0^t\Phi(t,s)\tilde{B}_0\tilde{B}_0^\mathrm{T}\Phi(t,s)^\mathrm{T}\,ds$ is the solution to $\dot{Y}=\tilde{A}_tY+Y\tilde{A}_t^\mathrm{T}+\tilde{B}_0\tilde{B}_0^\mathrm{T}$. Provided $P_1>0$ and $P_2>0$, it is easy to verify that a proper solution has 
	$$\lim_{t\rightarrow\infty}Y(t)=\begin{bmatrix}
		P_2&0 \\ 0& P_1
	\end{bmatrix},$$
	which is followed by  
	\begin{equation}\label{equ_tracking_error_with_w0}
		\|e\|_\mathcal{P} = \sqrt{\mathrm{tr}\left(C(P_1+P_2)C^\mathrm{T} + D_0D_0^\mathrm{T}\right)}.
	\end{equation}

	If the non-noisy system is suffered from the disturbance $w_1$, we third introduce the controller (\ref{equ_tildeQ_controller}) using a similar procedure, except that $L_t$ is replaced by $L_k$, and tracking error is measured by (\ref{equ_tracking_error_with_w1}) according to Theorem \ref{thm_tildeQ_control}. 

	In summary, one can conclude that the tracking error is estimated by 
	\begin{enumerate}
		\item Eq.~(\ref{equ_tracking_error_with_w0}) if $r\neq 0$, $w_0\neq 0$, $w_1=0$,
		\item Eq.~(\ref{equ_tracking_error_with_w1}) if $r= 0$, $w_0= 0$, $w_1 \neq 0$. 
	\end{enumerate}
	Notice that $r$, $w_0$ and $w_1$ are independent signals, which, together with the definition of power norm, gives (\ref{equ_tracking_error_with_w0w1}) and completes this proof.
\end{pf}

\begin{rem}
	In the proof, it can be seen that undesirable factors are handled by different control loops, and if non-nominal items do not exist, the nominal performance appears. 
\end{rem}

\subsection{Linear Systems with Additional Modeled Disturbance}

\begin{figure}[!htp]
	\centering
	\includegraphics[width=8.3cm]{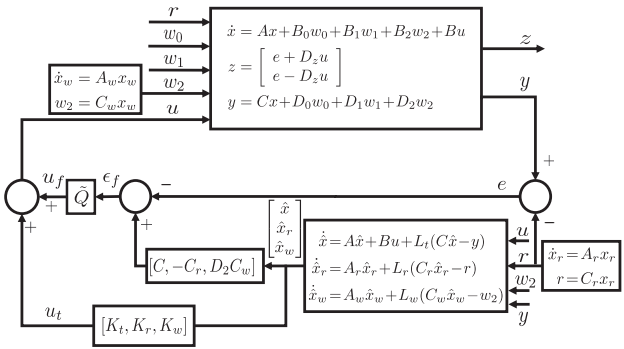}
	\caption{Control Structure for System with Modeled Disturbance}
	\label{fig_novel_structure_with_w2}
\end{figure}

For the case where partial disturbance is modeled, we denote $w_2\in\mathbb{R}^{m_2}$ as the modeled disturbance that is governed by (\ref{equ_disturbance_dynamics}), and the control structure is shown in Fig.~\ref{fig_novel_structure_with_w2}. 

The rejection of $w_2$ can be handled in the output regulation scheme (\ref{equ_OR_controller_hat_x}), (\ref{equ_OR_controller_hat_xr}) and 
\begin{align}
	\dot{\hat{x}}_w &=A_w\hat{x}_w+L_w(C_w\hat{x}_w-w_2), \label{equ_OR_controller_hat_xw} \\ 
	u_t&=K_t\hat{x}+K_r\hat{x}_r +K_w\hat{x}_w, \label{equ_OR_controller_u_2}
\end{align}
where $K_t$, $L_r$ and $L_k$ are the same as that in (\ref{equ_OR_controller_hat_xr}), (\ref{equ_OR_controller_u}) and (\ref{equ_OR_controller_Kalman}), $L_w$ is designed such that $(A_w+L_wC_w)$ is Hurwitz, and $K_r$ and $K_w$ satisfy (\ref{equ_OR_condition_1}), (\ref{equ_OR_condition_2}) and 
\begin{align*}
	Y_tA_w &= AY_t+BK_w+BK_tY_t+B_2C_w, \\ 
	0&=CY_t+D_2C_w,  
\end{align*}
for some matrices $X_t$ and $Y_t$. 

\begin{thm}\label{thm_control_with_w0w1w2}
	Given reference dynamics (\ref{equ_reference_dynamics}) satisfying Assumptions (A\ref{asm_OR_detectable_ArAw})--(A\ref{asm_OR_solvable_ArAw}), consider system (\ref{equ_system_with_w0w1w2}) satisfying Assumptions (A\ref{asm_stabilizable})--(A\ref{asm_A_B_column_rank}) under ontroller (\ref{equ_OR_controller_Kalman}), (\ref{equ_OR_controller_hat_xr}), (\ref{equ_OR_controller_hat_xw}), (\ref{equ_OR_controller_u_2}) and (\ref{equ_tildeQ_controller}). The tracking error $e$ satisfies (\ref{equ_tracking_error_with_w0w1}).
\end{thm}

\begin{pf}
	The proof of tracking performance and disturbance rejection is similar to that of Theorem \ref{thm_tildeQ_control} and \ref{thm_control_with_w0w1}. Thus it is omitted here. 
\end{pf}

\subsection{Linear System when $(C,A)$ is Not Detectable} 

\begin{figure}[!htp]
	\centering
	\includegraphics[width=8.3cm]{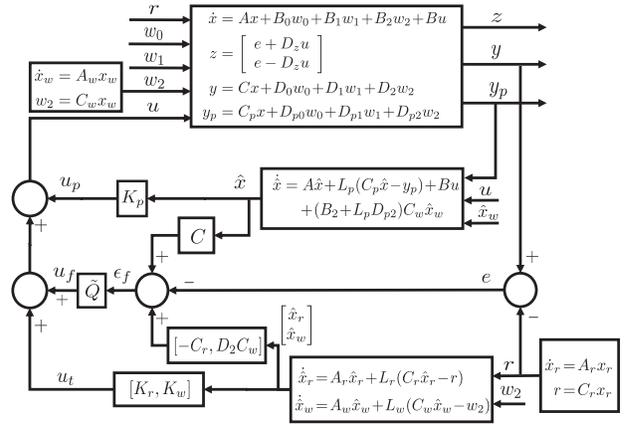}
	\caption{Three Loops Control Structure}
	\label{fig_control_structure_3loops}
\end{figure}

If $(C,A)$ is not fully detectable, it is reasonable to use all the output in the stabilization. In this subsection, we extend our control structure to a more general circumstance, where the output of the system is more than the reference dynamics by separating the stabilization out of the output regulation scheme. The structure of which is shown in Fig.~\ref{fig_control_structure_3loops}.

Consider the system 
\begin{equation}\label{equ_system_for_3loops}
	\begin{split}
		\dot{x} &= Ax+B_0w_0+B_1w_1+B_2w_2+Bu, \\
		y_p &= C_px+D_{p0}w_0+D_{p1}w_1+D_{p2}w_2, \\
		y &= Cx+D_0w_0+D_1w_1+D_2w_2, \\
		z &= \begin{bmatrix}
			e+D_zu \\
			e-D_zu
		\end{bmatrix}, \\
		e&=y-r, \\
		u &= u_p+u_t+u_f, 
	\end{split}
\end{equation}
where $y_p\in\mathbb{R}^q$ collects all measurable signals that can be used for stabilization, while $y$ contains the output that should track the reference $r$. $u_p$, $u_t$ and $u_f$ are designed for stabilization, output regulation and unmodeled disturbance rejection. Instead of Assumption (A\ref{asm_stabilizable}), we assume in the following that 
\begin{itemize}
	\setlength{\itemindent}{5.5mm}
	\item [(A\ref{asm_stabilizable})$^\prime$] 
		$(A,B)$ is stabilizable and $(C_p,A)$ is detectable.
\end{itemize}
Assumption (A\ref{asm_stabilizable})$^\prime$ decreases the conservatism of the Assumption (A\ref{asm_stabilizable}). Then we design the stabilization controller 
\begin{equation}\label{equ_controller_loop1}
	\begin{split}
		\dot{\hat{x}} &= A\hat{x}+Bu-L_p(y_p-C_p\hat{x})+(B_2+L_pD_{p2})C_w\hat{x}_w, \\
		u_p &= K_p\hat{x},
	\end{split}
\end{equation}
the output regulation controller (\ref{equ_OR_controller_hat_xr}), (\ref{equ_OR_controller_hat_xw}) and 
\begin{equation}\label{equ_controller_loop2}
	u_t=K_r\hat{x}_r+K_w\hat{x}_w,
\end{equation}
and the disturbance rejection controller $\tilde{Q}$ designed for 
\begin{align*}
	\dot{\tilde{x}} &= \begin{bmatrix}
		A+BK_p & -L_pC_p \\
		0 & A+L_pC_p
	\end{bmatrix}\tilde{x} + \begin{bmatrix}
		-L_pD_{p0} \\
		B_0+L_pD_{p0}
	\end{bmatrix} w_0 \\
	&\quad + \begin{bmatrix}
		-L_pD_{p1} \\
		B_1+L_pD_{p1}
	\end{bmatrix} w_1+\begin{bmatrix}
		B \\
		0 
	\end{bmatrix}u_f  \\
	&\quad +\begin{bmatrix}
		-BK_r & -BK_w-(B_2+L_pD_{p2})C_w \\
		0 & (B_2+L_pD_{p2})C_w
	\end{bmatrix}\overline{w}_1, \\
	z&=\begin{bmatrix}
		C & C \\
		C & C
	\end{bmatrix}\tilde{x} + \begin{bmatrix}
		D_0 \\ D_0
	\end{bmatrix}w_0 + \begin{bmatrix}
		D_1 \\ D_1
	\end{bmatrix}w_1 + \begin{bmatrix}
		D_z \\ -D_z
	\end{bmatrix} u_f, \\
	\epsilon_f &= \begin{bmatrix}
		0 & -C
	\end{bmatrix}\tilde{x}-D_0w_0 - D_1w_1+\begin{bmatrix}
		CX_t & CY_t
	\end{bmatrix}\overline{w}_1,
\end{align*}
where $\overline{w}_1=\left[(x_r-\hat{x}_r)^\mathrm{T},(x_w-\hat{x}_w)^\mathrm{T}\right]^\mathrm{T}$, $L_p=-(B_0D_{p0}^\mathrm{T}+P_3C^\mathrm{T})R_p^{-1}$, $R_p=D_{p0}D_{p0}^\mathrm{T}$, $P_3>0$ and $P_4>0$ satisfy 
\begin{align*}
	&(A+L_pC_p)P_3+P_3(A+L_pC_p)^\mathrm{T} \\
	&\quad +(B+L_pD_{p0})(B+L_pD_{p0})^\mathrm{T}=0, \\
	& (A+BK_p)P_4+P_4(A+BK_p)^\mathrm{T}+L_pD_{p0}D_{p0}^\mathrm{T}L_p^\mathrm{T}=0,
\end{align*}
and $K_p$ meets
\begin{enumerate}
	\item $A+BK_p$ is a Hurwitz matrix;
	\item $A+BK_p+L_pC_p-\lambda I$ has full row rank for all eigenvalues $\lambda$ of $A_r$ and $A_w$.
\end{enumerate}

\begin{thm}\label{thm_controller_3loops}
	Given reference dynamics (\ref{equ_reference_dynamics}) satisfying Assumptions (A\ref{asm_OR_detectable_ArAw})--(A\ref{asm_OR_solvable_ArAw}). 
	Consider system (\ref{equ_system_for_3loops}) under controller (\ref{equ_controller_loop1}), (\ref{equ_controller_loop2}), together with $\mathcal{H}_\infty$ controller $\tilde{Q}$. Suppose that Assumptions (A\ref{asm_stabilizable})$^\prime$, (A\ref{asm_A_B0_column_rank})--(A\ref{asm_A_B_column_rank}) hold. The tracking error $e$ satisfies 
	\begin{equation*}
		\|e\|_\mathcal{P} = \sqrt{\mathrm{tr}\left(C(P_3+P_4)C^\mathrm{T} + D_0D_0^\mathrm{T}\right) + \gamma^2 \|w_1\|_\mathcal{P}^2}.
	\end{equation*}
\end{thm}

\begin{pf}
	Using similar skills in the proof of Theorem \ref{thm_tildeQ_control} and \ref{thm_control_with_w0w1}, one can easily get this conclusion.
\end{pf}

\section{An Illustrative Example}\label{sec_5}

\begin{figure}[!ht]
	\centering
	\includegraphics[width=8.3cm]{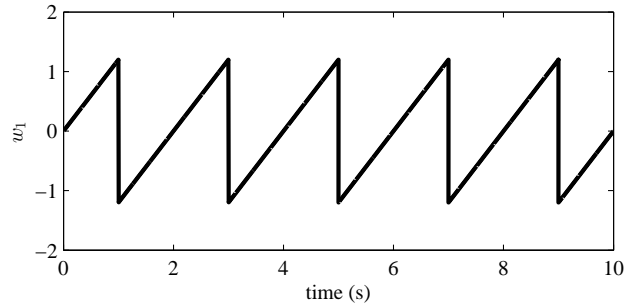}
	\caption{Unmodeled Disturbance}
	\label{fig_eg1_w1}
\end{figure}

In this section, our proposed robust tracking control framework is used in Furuta Inverted Pendulum, which is reported in \cite{cazzolato_dynamics_2011}. The system dynamics is governed by (\ref{equ_system_for_3loops}), where
$$A=\begin{bmatrix}
	0 &        0 &   1.0000 &        0 \\
	0 &        0 &        0 &   1.0000 \\
	0 & 149.2673 &  -7.0611 &  -0.9829 \\
	0 & 523.1909 &  -6.9788 &  -1.7226
\end{bmatrix},\;B=\begin{bmatrix}
	0 \\
	0 \\
  49.7260 \\
  49.1467 \\
  \end{bmatrix},$$
$$B_0=\left[\begin{array}{cccc}0.012 & 0 & 0 & 0 \\ 0 & 0.012 & 0 & 0 \\ 0 & 0 & 1 & 0 \\ 0 & 0 & 0 & 1\end{array}\right],\;
B_1=\left[\begin{array}{cccc}0.05 & 0 & 0 & 0 \\ 0 & 0.05 & 0 & 0 \\ 0 & 0 & 0.8 & 0 \\ 0 & 0 & 0 & 0.8\end{array}\right],$$
$$C_p=\begin{bmatrix}
	1 & 0 & 0 & 0 \\
	0 & 1 & 0 & 0
\end{bmatrix}, C=\left[1,\  0,\  0,\  0 \right], $$
$$D_{p0}=D_{p1} = \begin{bmatrix}
	0 & 0 & 1 & 0 \\
	0 & 0 & 0 & 1
\end{bmatrix}\times 10^{-4},$$
$$D_0=D_1=\left[0\;0\;1\;0\right]\times 10^{-4},$$
and state variable $x=\left[\theta_1,\theta_2,\dot{\theta}_1,\dot{\theta}_2\right]$ represents the angles of jointed arms and the angular velocities. The objective is to control the angle of the horizontal arm to track the reference signal $r=\sin(0.5\pi t)$ in radian, which is presented by the dynamics
\begin{equation*}
	\begin{split}
		\dot{x}_r&=\begin{bmatrix}
			0 & 1 \\ -(0.5\pi)^2 & 0
		\end{bmatrix}x_r,\quad x_r(0)=\begin{bmatrix}
			0 \\ 0.5\pi
		\end{bmatrix}, \\
		r &= \begin{bmatrix}
			1 & 0
		\end{bmatrix}x_r,
	\end{split}
\end{equation*}
in the presence of white noise, unmodeled disturbance as shown in Fig.~\ref{fig_eg1_w1}, and modeled disturbance governed by $w_2(t) = \cos(2\pi t)+0.5\cos(5\pi t)$. 

\begin{figure}[!ht]
	\centering
	\includegraphics[width=8.3cm]{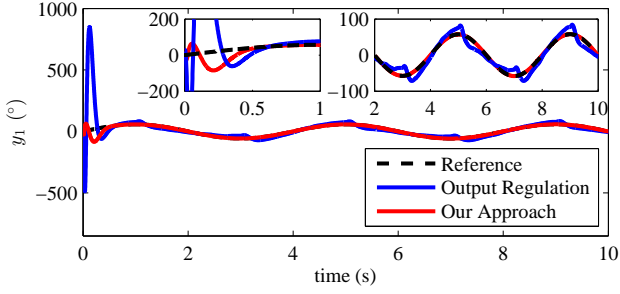}
	\caption{Angle of the horizontal arm}
	\label{fig_eg1_state_y1}
\end{figure}

\begin{figure}[!ht]
	\centering
	\includegraphics[width=8.3cm]{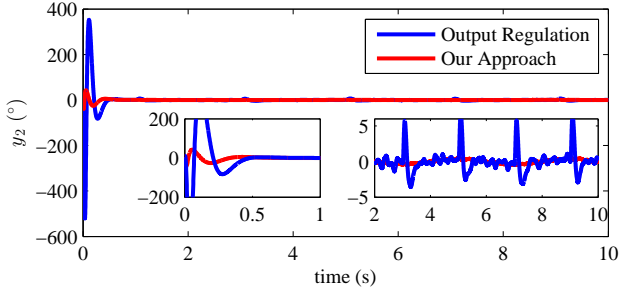}
	\caption{Angle of the vertical arm}
	\label{fig_eg1_state_y2}
\end{figure}

\begin{figure}[!ht]
	\centering
	\includegraphics[width=8.3cm]{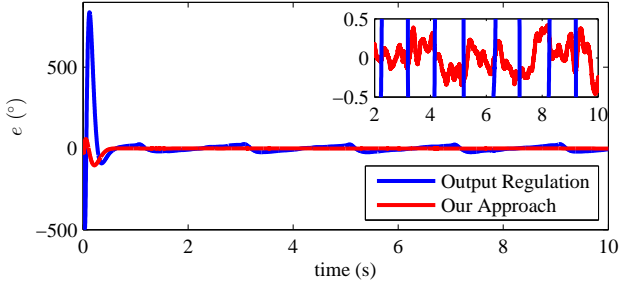}
	\caption{Tracking error}
	\label{fig_eg1_error}
\end{figure}

We will reveal the performance of our proposed control framework by comparison with output regulation control that is reported in \cite{huang2004nonlinear}. In the controller design, the angles of both arms are used in the stabilization, and the LMI based approach that was reported in \cite{Chilali1996LMIPolePlace} is adopted to assign poles into region $\mathcal{S}$ that is defined by
\begin{align*}
	\mathcal{S}(q,\theta,r)=\Big\{z:&\Re{z}<q,\ |\Im{z}|<\tan\theta\, |\Re{z}|,\ |z-q|<r,\\
	& q<0,\ 0\leq \theta<\frac{\pi}{2},\ r>0\Big\}.
\end{align*}
Specifically, the poles of $A+BK_p$ locates in $\mathcal{S}(-15,0.5,15)$ and $A_r+L_rC_r$ and $A_w+L_rC_w$ in $\mathcal{S}(-13,0.5,13)$. Then the controller reported in \cite{huang2004nonlinear} can be designed as 
$$K_p=\begin{bmatrix}
	4.7900 \\ -51.2190 \\ 1.2218 \\ -2.4165
\end{bmatrix}^\mathrm{T},\,
L_p=\begin{bmatrix}
	-0.0016 \\
	-0.0460 \\
	-0.1529 \\
	-1.0082 \\
\end{bmatrix}\times 10^5,$$
$$K_r=\begin{bmatrix}
	-4.5287 &  -1.0635
\end{bmatrix},$$
$$L_r=\begin{bmatrix}
	-39.8623 & -421.1467
\end{bmatrix}^\mathrm{T}, $$
$$K_w=\begin{bmatrix}
	-0.0891&  -0.0013&   -0.0878&  -0.0012
\end{bmatrix},$$
$$L_w=\begin{bmatrix}
	0.4801 & -0.1602 & -0.1600 & -0.1599 \\
	2.1791 & -0.7273 & -0.7263 & -0.7255 \\
	-0.3755 &  0.1253 &  0.1251 &  0.1250 \\
	3.3876 & -1.1303 & -1.1291 & -1.1284
\end{bmatrix}\times 10^7. $$

Let $\gamma = 0.34$ and $D_z=0.001$. According to Theorem \ref{thm_output_regulation} and \ref{thm_controller_3loops} the gains of our proposed control strategy can be designed as
$$L_p = \begin{bmatrix}
	-0.1134 &   0.0035 \\
	 0.0035 &  -0.1187 \\
	-9.2327 &   0.3394 \\
	 0.4767 &  -9.8542
 \end{bmatrix}\times 10^3,$$
$$K_f=\begin{bmatrix}
 -2.5832 &  2.3056 & -2.0880 &  2.0983 &  1.4468 \\
  3.0666 & -2.5820 &  2.4753 & -2.4322 & -3.3498 \\
 -0.1524 &  0.1287 & -0.1230 &  0.1209 &  0.1590 \\
  0.1552 & -0.1307 &  0.1253 & -0.1231 & -0.1695 \\
 -2.5864 &  2.3072 & -2.0895 &  2.1000 &  1.4308 \\
  3.0756 & -2.5864 &  2.4805 & -2.4375 & -3.3547 \\
 -0.1524 &  0.1287 & -0.1231 &  0.1209 &  0.1601 \\
  0.1552 & -0.1307 &  0.1253 & -0.1231 & -0.1721
\end{bmatrix}^{\mathrm{T}},$$
$$L_f=\begin{bmatrix}
  2.8192 &  0 & -0.0243 \\
  0.2664 &  0 & -0.0019 \\
 -5.5855 &  0 & -1.0247 \\
 -1.2274 &  0 & -0.0236 \\
  0.0004 &  0 & -0.0381 \\
  0.0033 &  0 & -0.0004 \\
  0.1629 &  0 &  0.3987 \\
  0.2038 &  0 &  0.0935
\end{bmatrix},$$
and $K_p$, $K_r$, $L_r$, $K_w$, $L_w$ are the same as that in the method of \cite{huang2004nonlinear}.

The angle trajectories of horizontal and vertical arms of the Inverted Pendulum are depicted in Fig.~\ref{fig_eg1_state_y1} and \ref{fig_eg1_state_y2}, respectively. From Fig.~\ref{fig_eg1_state_y1} and \ref{fig_eg1_state_y2}, one can see that the horizontal arm can track the reference signal under our proposed control strategy, and the maximum deviation of the vertical arm from the equilibrium point is less than $43^\circ$. In contrast, the system under the control scheme in \cite{huang2004nonlinear} exhibits a large overshoot ($520^\circ$ of the vertical arm), leading to possible unstable performance. The tracking error shown in Fig.~\ref{fig_eg1_error} indicates that the controller $\tilde{Q}$ narrows down the error in the transient and steady state, and the maximum tracking error is less than $0.5^\circ$, while the tracking error under controller in \cite{huang2004nonlinear} is over $27^\circ$.

\section{Conclusion}\label{sec_6}

In this paper, we proposed a design framework to solve the tracking problem of the linear system subject to noise, model disturbance and unmodeled disturbance, where an $\mathcal{H}_\infty$ control scheme based control loop $\tilde{Q}$ is introduced into the output regulation approach. All the output can be used in a separated stabilization loop, where the Kalman filter is adopted. In contrast to the output regulation scheme, our proposed framework provides a good balance between the robust stabilization and tracking performance in the control of the Furuta Inverted Pendulum.

\bibliographystyle{unsrt}
\bibliography{reference}

\end{document}